\documentclass[floatfix, prl, twocolumn, showpacs, showkeys, preprintnumbers, superscriptaddress,nofootinbib]{revtex4-1}
\usepackage[utf8]{inputenc}
\usepackage[sort&compress]{natbib}
\usepackage[normalem]{ulem}
\usepackage{bm}
\usepackage{times}
\usepackage{amssymb,amsbsy,amsmath,amsfonts}
\usepackage{graphicx}
\usepackage{float}
\usepackage{color}
\usepackage{morefloats}
\usepackage{rotating}
\usepackage{srcltx}
\usepackage{slashed}
\usepackage{subfigure}
\usepackage{multirow}
\usepackage{verbatim}
\usepackage{hyperref}
\usepackage{tabularx}
\usepackage{adjustbox}
\usepackage[dvipsnames]{xcolor}
\usepackage{nicefrac}

\usepackage{hyperref}
\hypersetup{
	colorlinks	=true,
	urlcolor	=blue,
	linkcolor	=blue,
	citecolor	=blue,
	pdftitle	={},
	pdfauthor	={Jun-Xu Lu, Li-Sheng Geng et al.},
	pdfsubject	={$\bar K N$ interaction to one loop}
	}

\begin{document}

\preprint{JLAB-THY-22-3716 }

\title{
Cross-channel constraints on resonant antikaon-nucleon scattering
}

\author{Jun-Xu Lu}
\affiliation{School of Space and Environment, Beijing 102206, China}
\affiliation{School of Physics, Beihang University, Beijing 102206, China}

\author{Li-Sheng Geng}
\email[Corresponding author: ]{lisheng.geng@buaa.edu.cn}
\affiliation{Peng Huanwu Collaborative Center for Research and Education, Beihang University, Beijing 100191, China}
\affiliation{School of
Physics,  Beihang University, Beijing 102206, China}
\affiliation{Beijing Key Laboratory of Advanced Nuclear Materials and Physics, Beihang University, Beijing 102206, China }
\affiliation{School of Physics and Microelectronics, Zhengzhou University, Zhengzhou, Henan 450001, China }
\author{Michael Doering}
%\email{doring@gwu.edu}
\affiliation{Institute for Nuclear Studies and Department of Physics, The George Washington University,
Washington, DC 20052, USA}
\affiliation{Thomas Jefferson National Accelerator Facility, 12000 Jefferson Avenue, Newport News,VA, USA}
\author{Maxim Mai}
\affiliation{Helmholtz-Institut f\"ur Strahlen- und Kernphysik (Theorie) and Bethe Center for Theoretical Physics, Universit\"at Bonn, D--53115 Bonn, Germany}
\affiliation{Institute for Nuclear Studies and Department of Physics, The George Washington University,
Washington, DC 20052, USA}

\begin{abstract}
Chiral perturbation theory and its unitarized versions have played an important role in our understanding of the low-energy strong interaction. Yet, so far, such studies typically deal  exclusively with perturbative or nonperturbative channels. In this letter, we report on the first  global study of meson-baryon scattering up to one-loop order. It is shown that covariant  baryon chiral perturbation theory, including its unitarization for the negative strangeness sector, can describe meson-baryon scattering data remarkably well. This provides a highly non-trivial check on the validity of this important low-energy effective field theory of QCD. We show that the $\bar{K}N$ related quantities can be better described in comparison with those of lower-order studies, and with reduced uncertainties due to the stringent constraints from the $\pi N$ and $K N$ phase shifts. In particular, we find that the two-pole structure of  $\Lambda(1405)$ persists up to one-loop order reinforcing the existence of two-pole structures in dynamically generated states. 

\end{abstract}

%\pacs{13.75.Gx, 13.75.Jz,12.39.Fe}

\maketitle

%%%%%%%%%%%%%%%%%%%%%%%%%%%%%%%%%
{\it Introduction:}
%%%%%%%%%%%%%%%%%%%%%%%%%%%%%%%%%
Resolving the patterns of low-energy hadron-hadron interactions  constitutes one of the most important goals of  modern theoretical physics, fostering not only a better understanding of the nonperturbative nature of QCD but also motivating novel tests of fundamental symmetries including searches for beyond standard model physics. Furthermore, the quantitative understanding of hadron-hadron interactions in the different strangeness sectors has important implications for matter on the largest scales in the universe. For example, in the strangeness zero sector, $\pi N$ scattering is related to the $\sigma_{\pi N}$ term, crucial for dark matter direct-detection efforts~\cite{Hill:2014yxa,Bottino:1999ei}. Another example with negative strangeness relates antikaon-nucleon scattering to the properties of neutron stars, where matter compressed to multiples of nuclear matter densities may allow for the appearance of kaon condensates~\cite{Gal:2016boi, Koch:1994mj}. Related to this is an ongoing controversy about the existence of deeply bound $K^-$ nuclear clusters~\cite{Batty:1997zp, Akaishi:2002bg, Magas:2006fn, Oset:2005sn,Yamazaki:2008hm, Maggiora:2009gh, Yamazaki:2010mu}, sensitive to the $\bar KN$ interaction. 
The interest in the strangeness sector has motivated several ongoing or future experiments including SIDDHARTA-2~\cite{Miliucci:2021drx,Curceanu:2020vjj,Zmeskal:2019ksw}, AMADEUS~\cite{AMADEUS:2021wln, Piscicchia:2018rez}, BGO-OD ~\cite{BGOOD:2021sog}, J-PARC E15, E31, E57 and E62~\cite{Hashimoto:2019qfy,Zmeskal:2015efj,J-PARCE15:2016esq,J-PARCE15:2020gbh}, PANDA~\cite{PANDA:2009yku,PANDA:2021ozp}, and the proposed  secondary $K_L$ beam in Hall D of Jlab~\cite{KLF:2020gai,Dobbs:2022agy}.

Bridging the different strangeness sectors of the meson-baryon interaction in a systematic fashion has been a major challenge, which is faced in this letter.
Systematic theoretical approaches to such interactions are provided  by lattice QCD and chiral perturbation theory (CHPT). Following the latter methodology\footnote{For reviews on recent progress in extracting resonant hadron-hadron dynamics from lattice QCD see, e.g., Refs.~\cite{Briceno:2017max,Mai:2022eur}.} baryon chiral perturbation theory (BCHPT) is known to be able to describe both pion-nucleon~\cite{Fettes:1998ud,Fettes:2000xg,Becher:2001hv,Mai:2009ce,Alarcon:2011zs,Chen:2012nx,Huang:2020iai} and kaon-nucleon~\cite{Lu:2018zof,Huang:2019not} scattering data rather well up to one-loop order. 
However,  in a strict perturbative application of BCHPT, such calculations do not allow to simultaneously study the antikaon-nucleon channel due to the existence of the subthreshold $\Lambda(1405)$-resonance. 
In this, unitarized CHPT with kernels from both leading~\cite{Kaiser:1995eg,Oset:1997it,Oller:2000fj,GarciaRecio:2002td} and next-to-leading order (NLO)~\cite{Borasoy:2006sr,Ikeda:2012au,Guo:2012vv,Mai:2012dt,Ramos:2016odk, Sadasivan:2022srs} CHPT became the predominant tool, including a prediction of a second pole~\cite{Oller:2000fj,Meissner:2020khl}. See Refs.~\cite{Meissner:2020khl,Mai:2020ltx,Hyodo:2020czb,Hyodo:2022xhp} for recent reviews.
Such models have also been shown to be consistent with modern experimental data, such as $K^- p\rightarrow \pi^0\pi^0\Sigma^0$ and $\gamma p\rightarrow K^+(\pi\Sigma)$ ~\cite{Magas:2005vu,Roca:2013cca,Mai:2014xna,Roca:2013av,Nakamura:2013boa}. Nevertheless, on a quantitative level, various NLO studies obtain different results depending on the details of the implementation~\cite{Cieply:2016jby,Bruns:2022fwz,Mai:2020ltx}, particular for the pole positions of the $\Lambda(1405)$. One hypothesis for the origin of this ambiguity is related to the large number of unknown low-energy constants (LECs) appearing in the NLO kernel. In this letter, we aim to improve on this by employing novel constraints from SU(3)-flavor symmetry and its breaking, simultaneously addressing the $\pi N$, $KN$ and $\bar KN$ channels.

A simultaneous study of all the meson-baryon scattering data is difficult for a number of reasons. First, it is well-known that 
an adequate description of  pion-nucleon scattering at energies below the first resonance ($\Delta(1232)$), i.e., $\sqrt{s}\approx1.16\,{\rm GeV}$, requires the inclusion of the next-to-next-to-leading order (NNLO) contributions~\cite{Lu:2018zof,Mai:2009ce}.
Second, the convergence of BCHPT in the three-flavor sector is controversial. Only in recent years it has been shown that this problem can be circumvented or  alleviated in the extended-on-mass-shell (EOMS) formulation of BCHPT~\cite{Gegelia:1999gf,Fuchs:2003qc}. For some recent applications, see, e.g.,  Ref.~\cite{Geng:2008mf} for the case of baryon magnetic moments, Ref.~\cite{Ren:2012aj} for the case of baryon masses, and Ref.~\cite{Geng:2013xn} for a  review. 

Of particular importance for the present work is the unifying NNLO BCHPT calculation of $\pi N$ and $K N$ scattering performed in Ref.~\cite{Lu:2018zof}. Unifying this with a nonperturbative formulation of the $\bar K N$-amplitude is the main challenge attacked in this work, which allows one to reliably determine the free parameters (more than 30 LECs) making advantage of SU(3) flavor symmetry and abundant scattering data. 

Notably, including $\bar K N$ data allows one to disentangle all LECs, instead of determining only certain combinations. In addition, for the $\bar KN$ channel, there is \emph{less} freedom in the NNLO fits with $KN$ and $\pi N$ constraints than in the NLO fits without them, which is  reflected in the reduced uncertainty. Indeed, apart from providing the first NNLO study of the $\bar KN$ sector, this paper also estimates, for the first time, systematic uncertainties for the meson-baryon sector from the truncation of the chiral expansion.

~\\
%%%%%%%%%%%%%%%%%%%%%%%%%%%%%%%%%
{\it Formalism:} 
%%%%%%%%%%%%%%%%%%%%%%%%%%%%%%%%%
In the following, we outline the workflow connecting the three meson-baryon channels $\pi N$, $KN$ and $\bar K N$. Beginning with the low-energy regime, we use BCHPT in the EOMS formulation~\cite{Gegelia:1999gf,Fuchs:2003qc} to write the scattering amplitude as
%%%%%%%%%%%%%%
\begin{align}
\label{ChiralAmplNNLO}
  T^{l\pm}_{\rm BCHPT}(s)=\mathcal{T}^{(1),l\pm}(s)+\mathcal{T}^{(2),l\pm}(s)+\mathcal{T}^{(3),l\pm}(s)\,
\end{align}
%%%%%%%%%%%%%%
up to the NNLO with respect to small meson momenta and masses, where the superscript $l\pm$ denotes that the amplitudes are projected to the partial waves with orbital angular moment $l$ and total angular momentum $J=l\pm\frac{1}{2}$. The total energy-squared is denoted by $s$. The terms $\mathcal{T}^{(i)}$ can be found in Ref.~\cite{Lu:2018zof}. As shown in the latter reference such expressions indeed allow to address the perturbative regime of the meson-baryon scattering below the lowest resonances quite successfully. Specifically, this is the case for the $KN$ and $\pi N$ channels close to respective thresholds. The situation is entirely different in the $S=-1$ channel, where the $\Lambda(1405)$-resonance appears around the $\bar KN$ threshold. Addressing this issue is guided by restoring two-body unitarity -- the fundamental principle of S-matrix theory -- for which we follow the method proposed in Ref.~\cite{Oller:2000fj}
%\com{mm}{$W\to\sqrt{s}$, $g\to G$}
%%%%%%%%%%%%%%
\begin{align}
\label{UnitaryAmp}
  &T^{l\pm}_{\bar K N}(s)=\left [1+N^{l\pm}(s)\cdot G(s)\right]^{-1}\cdot N^{l\pm}(s)\,,\\
  &N^{l\pm}(s)=T^{l\pm}_{\rm BCHPT}(s)+\mathcal{T}^{(1),l\pm}(s)\cdot G(s)\cdot\mathcal{T}^{(1),l\pm}(s)\,.\nonumber
\end{align}
%%%%%%%%%%%%%%
Here, all elements are promoted to matrices in channel space, ($\pi^0\Lambda$, $\pi^-\Sigma^+$, $\pi^0\Sigma^0$, $\pi^+\Sigma^-$, $K^-p$, $\bar{K}^0n$, $\eta\Lambda$,  $\eta\Sigma^0$, $K^0\Xi^0$, $K^+\Xi^-$), while $G(s)$ denotes a loop function, given explicitly in the Supplemental Material. 

%%%%%%%%%%%%%%%%%%%%%%%%
%%%%%%%%%%%%%%%%%%%%%%%%
\begin{table}[t]
\centering
\begin{tabular}{p{0.2\linewidth}p{0.18\linewidth}p{0.18\linewidth}p{0.18\linewidth}p{0.18\linewidth}}
\hline\hline
&$\bar{K}N$ &$\pi N$&$KN_{I=0}$&$KN_{I=1}$\\
\hline
$\chi^2/\mathrm{d.o.f.}$ &1.56 &1.28&0.46&1.46 \\
Data & 173 &78& 60& 60\\
\hline\hline
\end{tabular}
\caption{Best $\chi^2/\mathrm{d.o.f.}$ obtained for the four
isospin-strangeness meson-baryon channels. For the selection of the
$\pi N$ and $K N$ data, see Ref.~\cite{Lu:2018zof}. Note that the $\chi^2$'s for $\pi N$ and $KN$ defined in Ref.~\cite{Lu:2018zof} do not have statistical rigor and only serve to show the goodness of the fits.\label{chisquare}}
\end{table}
%%%%%%%%%%%%%%%%%%%%%%%%
%%%%%%%%%%%%%%%%%%%%%%%%

Characteristic to all unitarization procedures is the introduction of certain model-dependence~\cite{Mai:2020ltx} being also related to the appearance of the new unknown parameters $\{a_{\pi\Lambda}, a_{\pi\Sigma}, a_{\bar{K}N}, a_{\eta\Lambda}, a_{\eta\Sigma}, a_{K\Xi}\}$, corresponding to each combination of multiplets of the aforementioned channel space. 
However, the crucial feature of the employed program lies in the fact that both amplitudes~(\ref{ChiralAmplNNLO},\ref{UnitaryAmp}) coincide exactly when expanded to the third chiral order. Ultimately, this allows one to reduce uncertainties of extracted amplitudes and poles compared to the NLO calculations of individual strangeness sectors, by combining the analysis of perturbative ($\pi N,\,KN$) and nonperturbative sector ($\bar KN$) corresponding to Eqs.~\eqref{ChiralAmplNNLO} and \eqref{UnitaryAmp}. The larger number of fit parameters at NNLO compared to NLO is outweighed by the larger data base of the global analysis.

% %%%%%%%%%%%%%%%%%%%%%%%%
% %%%%%%%%%%%%%%%%%%%%%%%%
% \begin{table}[t]
% \centering
% \setlength{\tabcolsep}{0.7mm}{
% \begin{tabular}{ccc ccc ccc}
% \hline\hline
% $b_1^*$ &$b_2$&$b_3$&$b_4$&$b_5^*$&$b_6$&$b_7$&$b_8$&$c_1^*$ \\
% $-4.48$ &$-2.67$&$-2.87$&$1.30$&$0.44$&$0.29$&$0.79$&$0.03$&$-0.39$\\
% \hline
% $c_2$&$c_3$&$b_0^*$&$b_D^*$&$b_F^*$&$d_1$ &$d_2^*$&$d_3$&$d_4^*$\\
% $1.72$&$2.98$&$-0.62$&$0.06$&$-0.40$&$0.22$&$0.63$&$-1.39$&$3.18$\\
% \hline
% $d_5$&$d_6$&$d_7$ &$d_8^*$&$d_9$&$d_{10}$&$d_{48}$&$d_{49}^*$&$d_{50}$\\
% $1.01$&$-7.44$&$-2.64$&$-0.22$&$-5.95$&$1.67$&$-0.91$&$-0.10$&$0.84$\\
% \hline
% $a_{\pi\Lambda}$ &$a_{\pi\Sigma}$&$a_{\bar{K}N}$&$a_{\eta\Lambda}$& $a_{\eta\Sigma}$&$a_{K\Xi}$&& \\
% $3.94$ &$1.98$&$1.01$&$-0.33$&$3.26$&$-5.16$&& \\
% \hline\hline
% \end{tabular}
% }
% \caption{
% \label{para}
% Relevant meson-baryon LECs   up to NNLO ($b_0,b_D,b_F,b_{1,\cdots,4}$, $c_{1,2,3}$ in unit of GeV$^{-1}$, $b_{5,\cdots,8}$, $d_{4,5,6},d_{48,49,50}$ in units of GeV$^{-2}$, $d_{7,\cdots,10}$ in units of GeV$^{-3}$ and $d_{1,2,3}$ in units of GeV$^{-4}$) and the six substraction constants in the $\bar{K}N$ channel. Statistical uncertainties (percent level)  are not quoted, being sub leading to the systematic ones, e.g., originating from different fit strategies.}
% \end{table}
% %%%%%%%%%%%%%%%%%%%%%%%%
% %%%%%%%%%%%%%%%%%%%%%%%%

\begin{figure*}[t]
\centering
\includegraphics[width=1.0\textwidth]{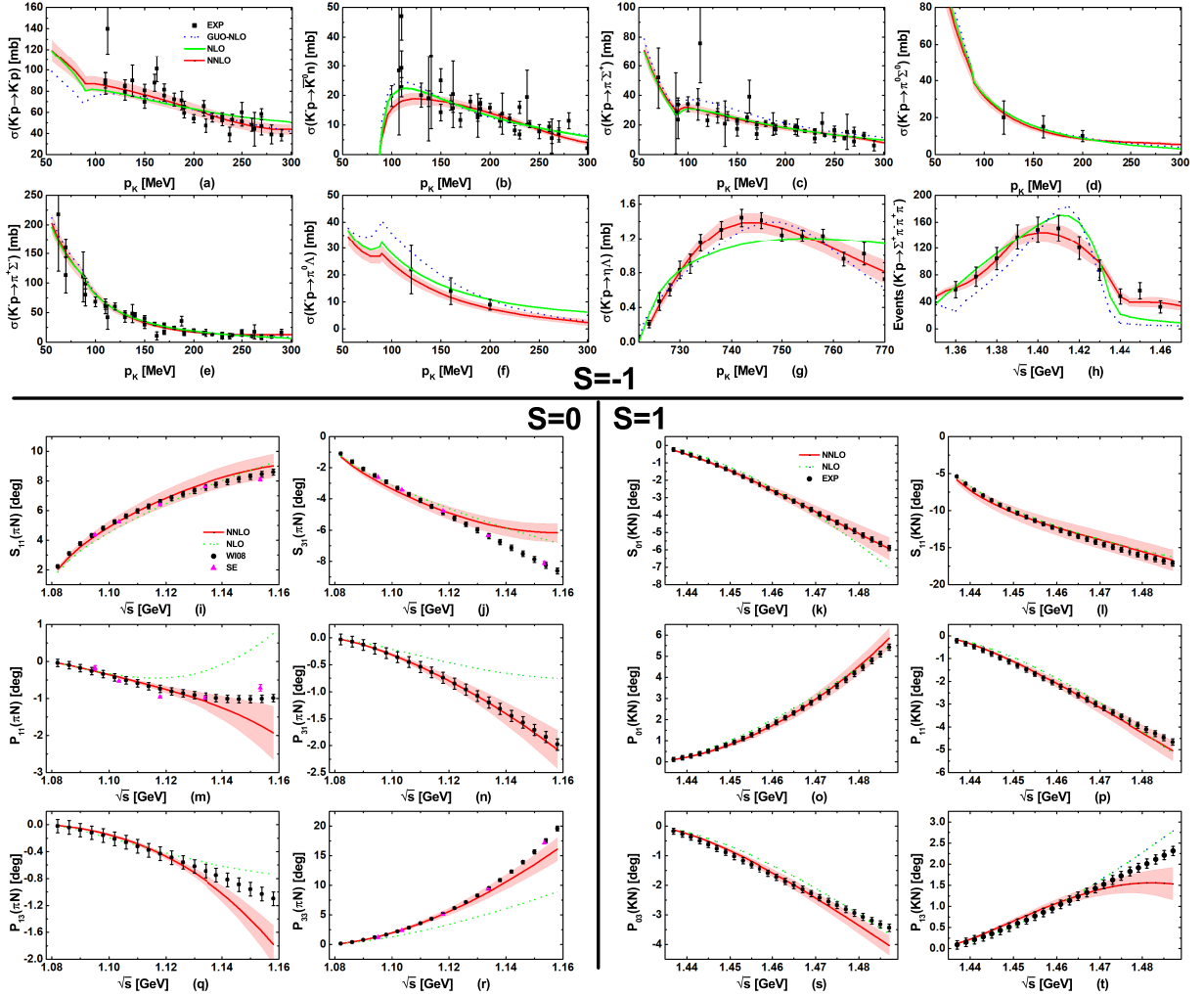}
\caption{Compilation of fit results. Upper panel: Cross sections for $\{K^-p\to X|X=K^-p,\bar{K}^0n,\pi^-\Sigma^+,\pi^0\Sigma^0,\pi^+\Sigma^-,\pi^0\Lambda,\eta\Lambda\}$ with $p_K$ denoting the $K^-$ laboratory momentum.The black dots with error bars denote the experimental data for total cross sections~\cite{Humphrey:1962zz,Kim:1965zzd,Sakitt:1965kh,Kittel:1966zz,Evans:1983hz,Ciborowski:1982et} and $\pi^-\Sigma^+$ mass spectrum ~\cite{Hemingway:1984pz}, while the red  bands (green dotted lines) refer to our NNLO (NLO) results, and the blue dashed lines denote the results of the NLO study~\cite{Guo:2012vv}.
Lower panel: Phase shifts of $\pi N$ and $K N$ elastic scattering compared to (black dots) SAID WI08~\cite{Arndt:2006bf} and SP92 solutions~\cite{Hyslop:1992cs}, respectively. The green dashed lines and red sold lines denote our NLO and NNLO results, respectively with uncertainty bands estimated by the Bayesian model. 
We also show the single energy(SE) $\pi N$ phase shifts (magenta triangles) obtained with a correlated analysis of different partial waves in Ref.~\cite{Doring:2016snk}.
}
\label{crosssection:KbarN}
\end{figure*}

~\\
%%%%%%%%%%%%%%%%%%%%%%%%%%%%%%%%%
{\it Results and discussions:}
%%%%%%%%%%%%%%%%%%%%%%%%%%%%%%%%% 
The formalism proposed in this letter is derived from the $\mathcal{O}(p^3)$ Chiral Lagrangian~\cite{Krause:1990xc,Oller:2006yh,Frink:2006hx}. The counterterms appearing in it contain 27 LECs, 14 of $\mathcal{O}(p^2)$ ($b_0, b_D, b_F, b_{1,\cdots,8}$ and $c_{1,2,3}$) and 13 of $\mathcal{O}(p^3)$ ($d_{1,\cdots,10,48,49,50}$). 
Taking into account  the six subtraction constants $\{a_{\pi\Lambda}, a_{\pi\Sigma}, a_{\bar{K}N}, a_{\eta\Lambda}, a_{\eta\Sigma}, a_{K\Xi}\}$ yields a total of 33 free parameters. Note that SU(3) flavor symmetry is broken both at the Lagrangian level and through the use of phycial hadron masses and channel specific subtraction constants.

The challenging task of determining them is approached in
a multi-step fitting strategy. In that, we note that $\pi N$ scattering data can be described well~\cite{Lu:2018zof} using $\mathcal{O}(p^3)$ amplitudes with eight combinations of the 27 LECs, see the Supplemental Material. Additionally, for each of the two $KN$ isospin channels, eight combinations of LECs contribute up to this order
which indeed decouple among all the three reaction channels $\{\pi N,KN_{I=0},KN_{I=1}\}$. Therefore, we 
leave the LECs related to the $\pi N$ 
scattering data and baryon masses fixed at the values determined in Ref.~\cite{Lu:2018zof}, denoting them with ``*'' in the corresponding tables in the Supplemental Material. 
For the remaining 23 free parameters, we first search for reasonable values for the 16 effective LECs which describe the $KN$ elastic scattering data. Then, taking these numbers as initial values, we perform a global fit for all the 23  parameters to the experimental data in the $\{KN,\bar KN\}$ sector.~\footnote{We also perform an alternative fit where the constraints from baryon masses are excluded, referred to as ``NNLO*'' afterwards. See the Supplemental Material for details.} Here, the latter consists of cross sections $\{K^-p\to X|X=K^-p,\bar{K}^0n,\pi^-\Sigma^+,\pi^0\Sigma^0,\pi^+\Sigma^-,\pi^0\Lambda,\eta\Lambda\}$~\cite{Humphrey:1962zz,Kim:1965zzd,Sakitt:1965kh,Kittel:1966zz,Evans:1983hz,Ciborowski:1982et} as well as threshold ratios $\{\gamma,R_c,R_n\}$~\cite{Nowak:1978au,Tovee:1971ga} and $K^-p$ scattering length ~\cite{Meissner:2004jr} extracted from the SIDDHARTA data~\cite{Bazzi:2011zj}.
Note that we only fit to the experimental data that are directly related to the two-body scattering amplitude $T_{M_iB_i\to M_jB_j}$ but do not consider the $\pi\Sigma$ spectra in the $\gamma p$~\cite{Ahn:2003mv,Niiyama:2008rt,CLAS:2013rjt}, $K^-p$~\cite{CrystallBall:2004ovf}, $pp$~\cite{Zychor:2007gf,HADES:2012csk}, and $e^- p$~\cite{CLAS:2013zie}  reactions, which involve at least three final-state particles. Theoretical efforts in this direction are being made currently, see, e.g., Refs.~\cite{Bruns:2022fwz,Mai:2014xna,Roca:2013av,Anisovich:2020lec}.

The best fit $\chi^2/\mathrm{d.o.f.}$ are listed in Table \ref{chisquare}. The corresponding LECs are shown in the Supplemental Material. 
They show that BCHPT and its unitarized version can provide a  good description of the meson-baryon scattering data for all the three strangeness sectors simultaneously. For the $\bar KN$ channel, with all the constraints from the $KN$ and $\pi N$ channels, we obtain a $\chi^2/\mathrm{d.o.f}=1.56$ weighting different observables by the respective number of data points~\cite{Hohler:1976ax,Borasoy:2006sr,Mai:2012dt,Guo:2012vv}, which should be compared with the equivalent value of about 2 from the NLO study~\cite{Guo:2012vv}. 
The $\chi^2/\mathrm{d.o.f.}$ for the $KN$ channels decrease considerably (from 3.93(2.24) to 0.46(1.46) for $KN_{I=0}$($KN_{I=1}$)) compared to those obtained in Ref.~\cite{Lu:2018zof} since we take into account the $\mathrm{O}(p^3)$ tree level contributions which were omitted there.

In Fig.~\ref{crosssection:KbarN}, we show the cross sections from the global NLO~\footnote{The NLO study is presented only for the sake of comparison. The description of  the $\bar{K}N$ channel is acceptable but that of the $\pi N$ channel is much worse. See the Supplemental Material for details.}  and NNLO fits for the $\bar{K}N$ coupled channels as well as $\pi N$ and $KN$ phase shifts. The error bands are produced by the Bayesian model for a degree of belief of 68\%~\cite{Furnstahl:2015rha,Melendez:2017phj,Melendez:2019izc} (see the Supplemental Material for details). The comparison with the best NLO fits of Guo~\cite{Guo:2012vv} reveals that the $\bar{K}N$ cross sections can be described rather well already at NLO, but quantitatively better results are obtained at NNLO, in particular, those of $\{\pi^-\Sigma^+, \pi^0\Lambda,\eta\Lambda\}$ final states. It is important to note  that compared to the NLO fits, only NNLO fits allow also for a simultaneous description of the $\pi N$ and $KN$ phase shifts~\cite{Lu:2018zof}. 

In Fig.~(\ref{crosssection:KbarN}h), we also show the $\pi^-\Sigma^+$ mass spectrum in the vicinity of $\Lambda(1405)$. As explained above, these data are not fitted. They are calculated following the approach of Refs.~\cite{Guo:2012vv,Oller:2000fj} but including the contributions from $\pi\Lambda$ and $\eta\Lambda$. The $\eta\Sigma$ and $K\Xi$ channels are neglected because they are too far away from the energy region of our interest. While we are faced with the well-known problem that the left-hand cuts overlap with the unitary cuts below $\bar{K}N$ threshold (see Supplemental Material for details), the data is indeed described well.

%%%%%%%%%%%%%%%%%%
%%%%%%%%%%%%%%%%%%
\begin{table*}
\centering
\setlength{\tabcolsep}{2.7mm}{
\begin{tabular}{p{0.15\linewidth}lll lll}
\hline\hline
&$a_{K^-p}$ [fm]  &  $\gamma$  &  $R_c$  & $R_n$    \\
\hline
NNLO                  & $(-0.71\pm0.07)+i(0.84\pm0.07)$      &  $2.35\pm0.19$    &  $0.684\pm0.033$ & $0.198\pm0.019$   \\
NLO~\cite{Guo:2012vv} &$-0.61^{+0.07}_{-0.08}+i(0.89^{+0.09}_{-0.08})$ & $2.36^{+0.17}_{-0.22}$ &  $0.661^{+0.12}_{-0.11}$ & $0.188^{+0.028}_{-0.029}$ \\
EXP                   &$(-0.64\pm0.10)+i(0.81\pm0.15)$  &  $2.36\pm0.12$  &  $0.664\pm0.033$ & $0.189\pm0.015$\\
\hline\\[-3mm]
    &Pole positions [MeV] & $|g_{\pi\Sigma}|$ [GeV]  & $|g_{\eta\Lambda}|$ [GeV] & $|g_{\bar{K}N}|$ [GeV] & $|g_{K\Xi}|$ [GeV]&\\
\hline
$\Lambda(1380)$ & $1392\pm8-i(102\pm15)$ & $6.40\pm0.10$ & $3.01\pm0.15$ & $2.31\pm0.10$ & $0.45\pm0.01$ &\\
$\Lambda(1405)$ & $1425\pm1-i(13\pm4)$ & $2.15\pm0.07$ & $5.45\pm0.24$ & $4.99\pm0.08$ & $0.58\pm0.02$ &\\
\hline
\hline
\end{tabular}
}
\caption{Threshold parameters, pole positions and couplings of the two $I=0$ states obtained in the present work in comparison with experimental data and the results of Ref.~\cite{Guo:2012vv}.
}\label{fitMAKN1}
\end{table*}
%%%%%%%%%%%%%%%%%%
%%%%%%%%%%%%%%%%%%

In Table~\ref{fitMAKN1} we compare the scattering length and three ratios with the experimental data.  Clearly the agreement is very good. We show as well the results of Fit II of the NLO study of Ref.~\cite{Guo:2012vv}, which  agree with ours within uncertainties. 

\begin{figure}[t]
\centering
\includegraphics[width=\linewidth, trim={1.5cm 0 3.5cm 1.9cm},clip]{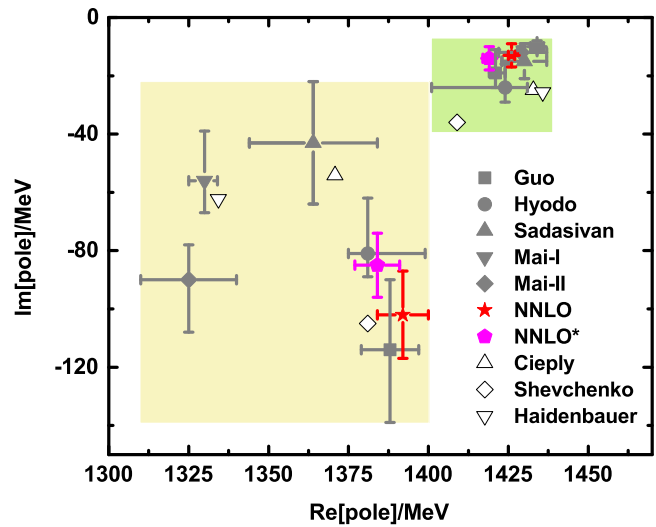}
\caption{Positions  of the two $\Lambda(1405)$ poles obtained in the present study (``NNLO'' and ``NNLO$^*$'' corresponding to results with or without baryon mass constraints) in comparison with those of the NLO studies, i.e., Guo~\cite{Guo:2012vv}, Hyodo~\cite{Ikeda:2012au},  Mai-I~\cite{Mai:2014xna},  Mai-II~\cite{Mai:2014xna}, Sadasivan~\cite{Sadasivan:2018jig}, Cieply~\cite{Cieply:2011nq}, Shevchenko~\cite{Shevchenko:2011ce}, Haidenbauer~\cite{Haidenbauer:2010ch}. 
}
\label{con:kn}
\end{figure}

The double pole structure of $\Lambda(1405)$ is the most interesting nonperturbative phenomenon in this coupled-channel problem. Studies on this special resonance date back to 1960s~\cite{Dalitz:1959dn} where it was suggested as a  $\bar{K}N$ bound state (see also review~\cite{Mai:2020ltx}). It was then found that $\Lambda(1405)$ is actually a superposition of two poles~\cite{Oller:2000fj,Oset:2001cn,Jido:2002yz,Jido:2003cb}. Recent discussions on this issue can be found in Refs.~\cite{Guo:2012vv,Ikeda:2011pi,Ikeda:2012au,Mai:2014xna,Sadasivan:2018jig,Bruns:2019bwg}. Note that a recent lattice QCD study also supports the $\bar{K}N$ bound state interpretation of $\Lambda(1405)$ ~\cite{Hall:2014uca}, see also Refs.~\cite{Molina:2015uqp, MartinezTorres:2012yi}. In order to obtain the pole position, one needs to extend the amplitudes to the second Riemann sheet. This can be achieved by analytically extrapolating the loop function $G(s)$ to the second Riemann sheet following the standard prescription, see, e.g., Refs.~\cite{Guo:2012vv,Mai:2022eur,Cieply:2016jby}. The poles discussed in the following are all situated on the respective sheet that is closest to the physical axis.
The couplings of the poles to various channels $i,j$ are obtained from the residues of the poles on the complex plane as $T^{ij}(s)=\lim_{s\rightarrow s_R}g_ig_j/(s-s_R)$.
%%%%%%%%%%%%
%\begin{equation}\label{cop}
%  T^{ij}(s)=\lim_{s\rightarrow s_R}\frac{g_ig_j}{s-s_R}\,.
%\end{equation}
%%%%%%%%%%%%
With the LECs determined above, we can predict the positions of the two poles and the corresponding couplings to various channels, which are shown in Table~\ref{fitMAKN1}. In the $I=0$ sector, the lower pole is located at $(1392,-102)$ MeV while the higher one at $(1425,-13)$ MeV. We also find a state located at $(1676,-25)$ MeV corresponding to the $\Lambda(1670)$-resonance. A selected compilation of the two-pole positions is shown in Fig.~\ref{con:kn} including the two-pole position from the NNLO and NNLO* fits corresponding to results with or without baryon mass constraints. It is clear that though the positions of the lower pole from different studies are quite scattered, those of the higher pole are determined much more precisely. We note that compared to the NLO results, the uncertainties in the NNLO results are  smaller, due to the  stringent constraints from the $\pi N$ and $K N$ scattering data. It is interesting to point out that the $\Lambda(1405)$ pole positions are similar to those of Fit II of Ref.~\cite{Guo:2012vv}.

~\\
%%%%%%%%%%%%%%%%%%%%%%%%%%%
{\it Conclusion and outlook:}
%%%%%%%%%%%%%%%%%%%%%%%%%%%
We have performed for the first time a global study of meson-baryon scattering in all three strangeness sectors $S=0,+1,-1$. The crucial step for this was the derivation of the formalism based on  covariant baryon chiral perturbation theory including next-to-next-to-leading order contributions while employing a consistent unitarization procedure for the nonperturbative $S=-1$ sector. Besides theoretical relevance, this formalism allows one to put tighter constraints on extracted amplitudes and resonances, by connecting data from the different reactions $\{\pi N,KN,\bar KN\}$, ensured by the $SU(3)_f$ symmetry and its breaking. Indeed, this is only possible at NNLO due to the known poor convergence of the chiral expansion in the $S=0$ sector.

Focusing on the $\bar KN$ sector, we confirmed the two-pole structure of $\Lambda(1405)$ in this novel approach, simultaneously ensuring for the first time an agreement with the perturbative channels. For the corresponding pole positions, we found  results consistent with most NLO studies but with  reduced uncertainties due to  the stringent constraints from the $\pi N$ and $KN$ scattering data. It should be stressed that for dynamically generated states, the existence of two-pole structures seems to be a common phenomenon~\cite{Meissner:2020khl}. Some recent examples that have attracted considerable attention include the $K_1(1270)$~\cite{Geng:2006yb} and $D_0^*(2300)$~\cite{Albaladejo:2016lbb}. The two-pole structure of $\Lambda(1405)$ can be understood by following trajectories on which symmetries of the  hadron-hadron interactions  are restored~\cite{Jido:2003cb,Roca:2005nm,Meissner:2020khl,Bruns:2021krp}. As a result, the emergence of a two-pole structure can be viewed as a strong evidence supporting the molecular nature of the state under investigation.

\begin{acknowledgments} 
JXL thanks Dr. Zhi-Hui Guo for useful discussions. This work is partly supported by the National Natural Science Foundation of China under Grant No.11735003, No.11975041,  and No. 11961141004, and the fundamental Research Funds for the Central Universities. JXL acknowledges support from the National Natural Science Foundation of China under Grant No.12105006 and China Postdoctoral Science Foundation under Grant No. 2021M690008.
MD and MM are supported by the U.S. Department of Energy, Office of Science, Office of Nuclear Physics under grant No. DE-SC0016582 and contract DE-AC05-06OR23177. This material is based upon work supported by the National Science Foundation under Grant No. PHY 2012289. The work of MM is furthermore supported by the Deutsche
Forschungsgemeinschaft (DFG, German Research Foundation) and the NSFC through the funds provided to the Sino-German Collaborative Research Center CRC 110 “Symmetries
and the Emergence of Structure in QCD” (DFG Project-ID 196253076 -
TRR 110, NSFC Grant No.~12070131001).
\end{acknowledgments}
%%%%%%%%%%%%%%%%%%%%%%%%%%%%%%%%%%
\bibliography{KbarN}
%%%%%%%%%%%%%%%%%%%%%%%%%%%%%%%%%%

\clearpage
%%%%%%%%%%%%%%%%%%%%%%%%%%%%%%%%%%
%%%%%%%%%%%%%%%%%%%%%%%%%%%%%%%%%%
\begin{widetext}
\section{Supplemental material}
In this supplemental material, we present some  details regarding the theoretical formulation, the Bayesian truncation uncertainties, the global next-to-leading order fit,  the description of the $\pi N$ and $KN$ scattering phase shifts, as well as an alternative fit without baryon mass constraints.

\subsection{Technical details}

In this subsection, we spell out certain quantities though well-known but necessary for clarity and self-consistence. The loop function $G(s)$ can be obtained via the once-subtracted dispersion relation,
\begin{align}\label{UnitaryAmp2}
    &16\pi^2G(s)_i=a_i(\mu)+\log\frac{m_i^2}{\mu^2}-x_+\log\frac{x_+-1}{x_+}-x_-\log\frac{x_--1}{x_-}\\
    \label{UnitaryAmp3}
    &x_\pm=\frac{s+M_i^2-m_i^2}{2s}\pm\frac{\sqrt{-4s(M_i^2-i0^+)+(s+M_i^2-m_i^2)^2}}{2s}
\end{align}
with $a_i$ the subtraction constants corresponding to the ten coupled channels in particle basis with strangeness $S=-1$, i.e., $\eta\Sigma^0,\pi^-\Sigma^+,\pi^0\Sigma^0,\pi^+\Sigma^-,\pi^0\Lambda,\eta\Lambda,K^-p,\bar{K}^0n,K^0\Xi^0,K^+\Xi^-$.
The cross sections  for  $M_iB_i\rightarrow M_jB_j$ read
%%%%%%%%%%%%%%
\begin{equation}\label{EQ:crosssection}
  \sigma_{ij}=\frac{1}{16\pi s}\frac{|\vec{p}_j|}{\vec{p}_i}|T_{M_iB_i\rightarrow M_jB_j}|^2\,,
\end{equation}
%%%%%%%%%%%%%%
where $\vec{p}_i$, $\vec{p}_j$ are the momenta of incoming and outgoing particles in the center of mass frame.
The three  threshold ratios are directly related to the scattering cross sections at threshold as
%%%%%%%%%%%%%%
\begin{equation}\label{Eq:ratios}
  \begin{split}
    \gamma=  \frac{\sigma(K^-p\rightarrow \pi^+\Sigma^-)}{\sigma(K^-p\rightarrow \pi^-\Sigma^+)}\,,\qquad 
    R_c=  \frac{\sigma(K^-p\rightarrow \mathrm{charged~particles})}{\sigma(K^-p\rightarrow \mathrm{all})}, 
    R_n=  \frac{\sigma(K^-p\rightarrow \pi^0\Lambda)}{\sigma(K^-p\rightarrow \mathrm{all~neutral~states})}. \\
  \end{split}
\end{equation}
%%%%%%%%%%%%%%
The $K^-p$ scattering length is defined as
%%%%%%%%%%%%%%
\begin{equation}\label{Eq:scatteringlength}
  a_{K^-p}= \frac{1}{8\pi \sqrt{s}}T_{K^-p\rightarrow K^-p}(s)|_{\sqrt{s}=m_{K^-}+m_p}.
\end{equation}
%%%%%%%%%%%%%%

In dealing with the $\pi \Sigma$ spectrum in Fig.~\ref{crosssection:KbarN}(h), we are faced with the well-known problem in such a coupled-channel formalism that the unitary cuts overlap with the unphysical subthreshold cuts from the crossed diagrams, which include both the LO crossed Born terms and NNLO crossed loop diagrams. Coupled channels with higher thresholds, i.e., $\eta\Sigma$, $\eta\Lambda$, and $K\Xi$ introduce these additional cuts well above the $\pi\Sigma$ thresholds and even not far from the $\bar{K}N$ thresholds. These unphysical cuts lead to logarithmic divergences and violate the unitarity in the $S$-wave amplitudes. However, these unphysical cuts are only artifacts of on-shell amplitudes and will not be present in a full theoretical calculation. In the present work, we follow the strategy of Ref.~\cite{Borasoy:2005ie} and eliminate the unphysical cuts by matching the contribution of the crossed diagrams to a constant real value below a certain invariant energy $\sqrt{s_0}$. As it was pointed out in Ref.~\cite{Borasoy:2005ie}, the $s_0$ is somehow arbitrary as long as it is not too close to the singularities due to the numerically small contributions from crossed diagrams. Note that the cross sections, scattering lengths and the three ratios are not affected at all because these unphysical cuts are located well below the $K^- p$ threshold.

%%%%%%%%%%%%%%%%%%%%%%
\subsection{Bayesian truncation uncertainties and the NLO fit}
%%%%%%%%%%%%%%%%%%%%%%

In order to estimate the truncation uncertainties of our NNLO results, we perform a global NLO fit following the same fitting strategy adopted in the NNLO study.  The results are shown in Fig.~\ref{crosssection:KbarN} of the main text and Table~\ref{SCOP2}. 

It is clear that up to NLO, the total cross sections of the $K^-p$-induced reactions can be described quite well, similar to the results of Ref.~\cite{Guo:2012vv}, except for the $K^-p \rightarrow \eta \Lambda$ cross section which keeps increasing in the higher momentum region. However, for the elastic scattering phase shifts, significant discrepancies appear in all the partial waves except for $S_{11}$. In addition, the constraints from the $\pi N$ and $K N$ phase shifts deteriorate the description of threshold parameters in the $S=-1$ sector, as shown in Table~\ref{SCOP2}. Indeed, the present NLO fit (including these phase shifts) cannot fit the $K^-p$ scattering length and threshold branching ratios as well as the NLO fit of Ref.~\cite{Guo:2012vv} that does not include these phase shifts. On the other hand, we still find two $I=0$ poles  as shown in Table~\ref{SCOP2}. Compared to the pole positions obtained with the NNLO potentials and the Fit II results of Ref.~\cite{Guo:2012vv}, the lower $\Lambda(1380)$ is a bit narrower and the higher $\Lambda(1405)
$ is a bit broader. 

In principle, the truncation uncertainties should be estimated starting from LO  as was done in, e.g., the studies of the nuclear force~\cite{Epelbaum:2019zqc,Lu:2021gsb}. However, as widely recognized  in previous studies, although the total $\bar{K}N$ cross sections  could be described reasonably well, the elastic $\pi N$ scattering phase shifts differ significantly from the WI08 result even not far away from the threshold in the SU(2)  covariant NLO ChPT~\cite{Siemens:2016hdi}. The description is even worse if one treats this issue non-relativistically. This is also one of the motivations for  a global study up to the one-loop order in the present work. As a consequence, the truncation uncertainties will be unphysically large if we consider the LO contribution. Therefore, we choose to set the reference scale $X_{\rm{ref}}$ in the Bayesian truncation estimation as
\begin{align}
    X_{\rm{ref}}=\text{Max}\{\frac{|X^{\rm{NLO}}|}{Q^2},\frac{|X^{\mathrm{NNLO}}-X^{\mathrm{NLO}}|}{Q^2} \},
\end{align}
where we take $Q=\frac{m_{ave}}{\Lambda_b}$ with $m_{ave}=0.370$ GeV the average mass of pseudoscalar mesons and $\Lambda_b=1.16$~GeV the average mass of octet baryons.  We note that for the breakdown scale one can also take the chiral symmetry breaking scale $\Lambda_\mathrm{ChPT}=4\pi f_\pi\approx1.2$ GeV.  The appearance of the $\Delta(1232)3/2^+$ in the $P_{33}$ partial wave of the $\pi N$ amplitude signals that the breakdown scale for this particular channel is $m_\Delta - m_N$, beyond which either non-perturbative treatment or explicit inclusion of $\Delta(1232)$ is required. As a result, we have limited our fits to the $\pi N$ phase-shifts in the threshold region, like many comparable studies~\cite{Alarcon:2011zs,Chen:2012nx,Huang:2019not,Siemens:2016hdi}. For other details of the Bayesian model, we refer to Refs.~\cite{Furnstahl:2015rha,Melendez:2017phj,Melendez:2019izc}.

\begin{table*}
\centering
\begin{tabular}{ccccccc}
\hline\hline
&$a_{K^-p}$ [fm]  &  $\gamma$  &  $R_c$  & $R_n$    \\
\hline
NLO                  & $-0.82+i0.83$      &  $2.33$    &  $0.670$ & $0.211$   \\
NLO~\cite{Guo:2012vv} &$-0.61^{+0.07}_{-0.08}+i0.89^{+0.09}_{-0.08}$ & $2.36^{+0.17}_{-0.22}$ &  $0.661^{+0.12}_{-0.11}$ & $0.188^{+0.028}_{-0.029}$ \\\
EXP                   &$(-0.64\pm0.10)+i(0.81\pm0.15)$  &  $2.36\pm0.12$  &  $0.664\pm0.033$ & $0.189\pm0.015$\\
\hline
    &Positions [MeV]  & $|g_{\pi\Sigma}|$ [GeV]  & $|g_{\eta\Lambda}|$ [GeV] & $|g_{\bar{K}N}|$ [GeV] & $|g_{K\Xi}|$ [GeV] &\\
\hline
$\Lambda(1380)$ & $1367\pm12-i(47\pm16)$ & $7.50\pm1.85$ & $4.69\pm2.14$ & $6.72\pm2.92$ & $2.01\pm0.85$ &\\
$\Lambda(1405)$ & $1426\pm17-i(26\pm12)$ & $5.09\pm2.50$ & $5.98\pm1.82$ & $7.17\pm2.89$ & $2.49\pm0.82$ &\\
\hline\hline
\end{tabular}
\caption{Threshold parameters, pole positions and couplings of the two $I=0$ states obtained at NLO. }\label{SCOP2}
\end{table*}

\subsection{Elastic $\pi N$ and $KN$ scattering}
%%%%%%%%%%%%%%%%%%%%%%%%
%%%%%%%%%%%%%%%%%%%%%%%%
\begin{table}[t]
\centering
\setlength{\tabcolsep}{0.7mm}{
% \begin{tabular}{ccc ccc ccc}
% \hline\hline
% $b_1^*$ &$b_2$&$b_3$&$b_4$&$b_5^*$&$b_6$&$b_7$&$b_8$&$c_1^*$ \\
% $-4.48$ &$-2.67$&$-2.87$&$1.30$&$0.44$&$0.29$&$0.79$&$0.03$&$-0.39$\\
% \hline
% $c_2$&$c_3$&$b_0^*$&$b_D^*$&$b_F^*$&$d_1$ &$d_2^*$&$d_3$&$d_4^*$\\
% $1.72$&$2.98$&$-0.62$&$0.06$&$-0.40$&$0.22$&$0.63$&$-1.39$&$3.18$\\
% \hline
% $d_5$&$d_6$&$d_7$ &$d_8^*$&$d_9$&$d_{10}$&$d_{48}$&$d_{49}^*$&$d_{50}$\\
% $1.01$&$-7.44$&$-2.64$&$-0.22$&$-5.95$&$1.67$&$-0.91$&$-0.10$&$0.84$\\
% \hline
% $a_{\pi\Lambda}$ &$a_{\pi\Sigma}$&$a_{\bar{K}N}$&$a_{\eta\Lambda}$& $a_{\eta\Sigma}$&$a_{K\Xi}$&& \\
% $3.94$ &$1.98$&$1.01$&$-0.33$&$3.26$&$-5.16$&& \\
% \hline\hline
% \end{tabular}
\begin{tabular}{ccccccccccccccccc}
\hline\hline
$b_1^*$ &$b_2$&$b_3$&$b_4$&$b_5^*$&$b_6$&$b_7$&$b_8$ &$c_1^*$ &$c_2$&$c_3$&$b_0^*$&$b_D^*$&$b_F^*$&& \\
\hline
$-4.48$ &$-2.67$&$-2.87$&$1.30$&$0.44$&$0.29$&$0.79$&$0.03$&$-0.39$ &$1.72$&$2.98$&$-0.62$&$0.06$&$-0.40$&& \\
$d_1$ &$d_2^*$&$d_3$&$d_4^*$&$d_5$&$d_6$&$d_7$ &$d_8^*$&$d_9$&$d_{10}$&$d_{48}$&$d_{49}^*$&$d_{50}$&& \\
$0.22$&$0.63$&$-1.39$&$3.18$&$1.01$&$-7.44$&$-2.64$&$-0.22$&$-5.95$&$1.67$&$-0.91$&$-0.10$&$0.84$&& \\
$a_{\pi\Lambda}$ &$a_{\pi\Sigma}$&$a_{\bar{K}N}$&$a_{\eta\Lambda}$& $a_{\eta\Sigma}$&$a_{K\Xi}$&& \\
$3.94$ &$1.98$&$1.01$&$-0.33$&$3.26$&$-5.16$&& \\
\hline\hline
\end{tabular}
}
\caption{
\label{para}
Relevant meson-baryon LECs   up to NNLO  (fit with baryon mass constraints). The LECs $b_0,b_D,b_F,b_{1,\cdots,4}$, $c_{1,2,3}$ are quoted in unit of GeV$^{-1}$, the $b_{5,\cdots,8}$, $d_{4,5,6},d_{48,49,50}$ in units of GeV$^{-2}$, and the $d_{7,\cdots,10}$ in units of GeV$^{-3}$ and $d_{1,2,3}$ in units of GeV$^{-4}$) and the six substraction constants in the $\bar{K}N$ channel. Statistical uncertainties (percent level)  are not quoted, being sub leading to the systematic ones, e.g., originating from different fit strategies.  Stars indicate LECS that are kept fixed from the fit to $\pi N$ 
scattering data and baryon masses determined in Ref.~\cite{Lu:2018zof}.}
\end{table}
%%%%%%%%%%%%%%%%%%%%%%%%
%%%%%%%%%%%%%%%%%%%%%%%%

Here, we briefly explain the  $\pi N$ and $K N$ elastic scattering phase shifts shown in Fig.~\ref{crosssection:KbarN} of the main text with the LECs given in Table~\ref{para}. Since we have fixed the LECs related to the $\pi N$ phase shifts, they are actually the same as the results obtained in Ref.~\cite{Lu:2018zof}, complemented with the truncation uncertainties estimated above. We note that the $\pi N$ phase shifts cannot be well described at NLO, particularly those of the $P_{11}$, $P_{31}$ and $P_{33}$ partial waves. All the phase shifts are described well at NNLO though in the higher energy region the $S_{31}$ phase shifts of $\pi N$ and those of $P_{13}$ of $K N$ are a  bit worse due to the constraints from baron masses. See Fig.~\ref{crosssection:KbarNNOMA} for the fits obtained without the constraints from baryon masses referred to as ``NNLO*''. Indeed, the description of the two mentioned partial waves improves considerably at higher energies.

Table~\ref{final-LEC} shows the combinations of LECs corresponding to $\pi N$, $KN_{I=0}$, $KN_{I=1}$ elastic scattering  from Ref.~\cite{Lu:2018zof} for  reference. We stress again that through the inclusion of the $\bar KN$ sector, all LECs are fully disentangled and one can determine their individual values.

\begin{table*}
\centering
\begin{tabular}{c|c|c}
\hline\hline
$\pi N$  &  $KN_{I=0}$  &  $KN_{I=1}$   \\
\hline
 $\alpha_1=b_1+b_2+b_3+2b_4$  &  $\beta_1=b_3-b_4$         &  $\gamma_1=b_1+b_2+b_4$    \\
 $\alpha_2=b_5+b_6+b_7+b_8$   &  $\beta_2=2b_6-b_8$        &  $\gamma_2=2b_5+2b_7+b_8$  \\
 $\alpha_3=c_1+c_2$           &  $\beta_3=4c_1+c_3$        &  $\gamma_3=4c_2+c_3$       \\
 $\alpha_4=2b_0+b_D+b_F$      &  $\beta_4=b_0-b_F$         &  $\gamma_4=b_0+b_D$        \\
 $\alpha_5=d_2$               &  $\beta_5=d_1+d_2+d_3$     &  $\gamma_5=d_1-d_2-d_3$    \\
 $\alpha_6=d_4$               &  $\beta_6=d_4+d_5+d_6$     &  $\gamma_6=d_4-d_5+d_6$    \\
 $\alpha_7=d_8+d_{10}$        &  $\beta_7=d_7-d_8+d_{10}$  &  $\gamma_7=d_7+d_8+d_{10}$  \\
 $\alpha_8=d_{49}$            &  $\beta_8=d_{48}+d_{49}+d_{50}$  &  $\gamma_8=d_{48}+d_{49}-d_{50}$  \\
\hline\hline
\end{tabular}
\caption{Independent (combinations of) LECs  contributing to $\pi N$ and $KN$ scattering. }\label{final-LEC}
\end{table*}

\subsection{Alternative fit strategies}

As pointed out in Refs.~\cite{Ren:2012aj,Lu:2018zof}, although the physical baryon masses can be reproduced accurately up to $\mathcal{O}(p^3)$, the lattice QCD baryon masses cannot be described satisfactorily at this order. The LECs actually contribute at different orders to the baryon masses than to the scattering observables. In Ref.~\cite{Lu:2018zof}, we found that with the constraints from baryon masses on $b_0$, $b_D$ and $b_F$, the description of $\pi N$ phase shifts, particularly those of the $S_{31}$ partial wave, is a bit worse. Given the fact that up to N$^3$LO, both the physical baryon masses and those of lattice QCD simulations, as well as the sigma terms can be described quite well, it is also interesting to see how the description of  meson-baryon scattering changes if one neglects the constraints on the LECs from the baryon masses.

\begin{table}[t]
\centering
% \caption{}
% \label{chisqureNOMA}
\begin{minipage}{0.28\linewidth}
    \begin{tabular}{lccccc}
    \hline\hline
     &$\bar{K}N$ &$\pi N$&$KN_{I=0}$&$KN_{I=1}$\\
    \hline
    $\chi^2/\mathrm{d.o.f.}$ &1.53 &0.15&0.64&0.32 \\
    Data & 173 &78& 60& 60\\
    \hline\hline\\[1.8cm]
    \end{tabular}
\end{minipage}
~~
\begin{minipage}{0.68\linewidth}
\setlength{\tabcolsep}{0mm}
    \begin{tabular}{lcccccc}
    \hline\hline
    &$a_{K^-p}$ [fm]  &  $\gamma$  &  $R_c$  & $R_n$    \\
    \hline
    NNLO                  & $-0.65\pm0.08+i(0.80\pm0.07)$      &  $2.33\pm0.28$    &  $0.693\pm0.09$ & $0.195\pm0.03$   \\
    NLO~\cite{Guo:2012vv} &$-0.61^{+0.07}_{-0.08}+i0.89^{+0.09}_{-0.08}$ & $2.36^{+0.17}_{-0.22}$ &  $0.661^{+0.12}_{-0.11}$ & $0.188^{+0.028}_{-0.029}$ \\\
    EXP                   &$(-0.64\pm0.10)+i(0.81\pm0.15)$  &  $2.36\pm0.12$  &  $0.664\pm0.033$ & $0.189\pm0.015$\\
    \hline
    &Positions [MeV]  & $|g_{\pi\Sigma}|$ [GeV]  & $|g_{\eta\Lambda}|$ [GeV] & $|g_{\bar{K}N}|$ [GeV] & $|g_{K\Xi}|$ [GeV] &\\
    \hline
    $\Lambda(1380)$ & $1384\pm7-i(85\pm11)$ & $3.26\pm0.11$ & $0.87\pm0.02$ & $2.04\pm0.11$ & $0.61\pm0.02$ &\\
    $\Lambda(1405)$ & $1419\pm2-i(14\pm4)$ & $3.24\pm0.17$ & $0.42\pm0.02$ & $6.01\pm0.12$ & $0.81\pm0.03$ &\\
    \hline
    \hline\\
    \end{tabular}
\end{minipage}
\setlength{\tabcolsep}{2.3mm}{
\begin{tabular}{ccccccccccccccccc}
\hline\hline
$b_1^*$ &$b_2$&$b_3$&$b_4$&$b_5^*$&$b_6$&$b_7$&$b_8$ &$c_1^*$ &$c_2$&$c_3$&$b_0^*$&$b_D$&$b_F$&& \\
\hline
$-5.36$ &$4.37$&$-9.14$&$1.25$&$0.35$&$1.59$&$-0.36$&$-0.15$&$-0.91$ &$2.25$&$1.22$&$0.74$&$4.34$&$-8.55$&& \\
$d_1$ &$d_2^*$&$d_3$&$d_4^*$&$d_5$&$d_6$&$d_7$ &$d_8^*$&$d_9$&$d_{10}$&$d_{48}$&$d_{49}^*$&$d_{50}$&& \\
$-0.32$&$0.61$&$-1.19$&$3.25$&$-4.96$&$-7.53$&$-2.85$&$-0.00$&$-5.56$&$1.46$&$9.80$&$-0.32$&$-2.02$&& \\
$a_{\pi\Lambda}$ &$a_{\pi\Sigma}$&$a_{\bar{K}N}$&$a_{\eta\Lambda}$& $a_{\eta\Sigma}$&$a_{K\Xi}$&& \\
$4.80$ &$1.11$&$0.90$&$3.01$&$1.38$&$3.06$&& \\
\hline\hline
\end{tabular}}
\caption{Compilation of ``NNLO*'' results (without constraints from baryon masses): (top left) Best $\chi^2/\mathrm{d.o.f.}$ obtained for the four isospin-strangeness meson-baryon channels for the fit without constraints from baryon masses. For the selection of the $\pi N$ and $K N$ data, see Ref.~\cite{Lu:2018zof}; (top right) Threshold parameters, pole positions and couplings of the two $I=0$ states obtained at NNLO without constraints from baryon masses; (bottom) Relevant meson-baryon LECs up to NNLO ($b_0,b_D,b_F,b_{1,\cdots,4}$, $c_{1,2,3}$ in units of GeV$^{-1}$, $b_{5,\cdots,8}$, $d_{4,5,6},d_{48,49,50}$ in units of GeV$^{-2}$, $d_{7,\cdots,10}$ in units of GeV$^{-3}$ and $d_{1,2,3}$ in units of GeV$^{-4}$) and the six substraction constants in the $\bar{K}N$ channel. }\label{paraNOMA}
\end{table}

We show in Table~\ref{paraNOMA} the $\chi^2/\mathrm{d.of.}$ and LECs of the NNLO fit without the constraints from the baryon masses referred to as ``NNLO*''. Clearly, the descriptions of $\pi N$ and $K N$ phase shifts improve significantly. We notice that the LECs change considerably and in particular the  LECs $b_0$, $b_D$, $b_F$  are somehow unnaturally large. We show the total cross sections for $\bar{K}N$ and  phase shifts for $\pi N$ and $K N$ and in Fig.~\ref{crosssection:KbarNNOMA}, as well as the threshold parameters and pole positions in Table~\ref{paraNOMA}. Only slight improvements are observed for the observables obtained with the nonperturbative $\bar{K} N$ amplitudes. We find again the double-pole structure of $\Lambda(1405)$ with slightly lower masses and narrower widths.

\begin{figure*}[htpb]
\centering
\includegraphics[width=1.0\textwidth]{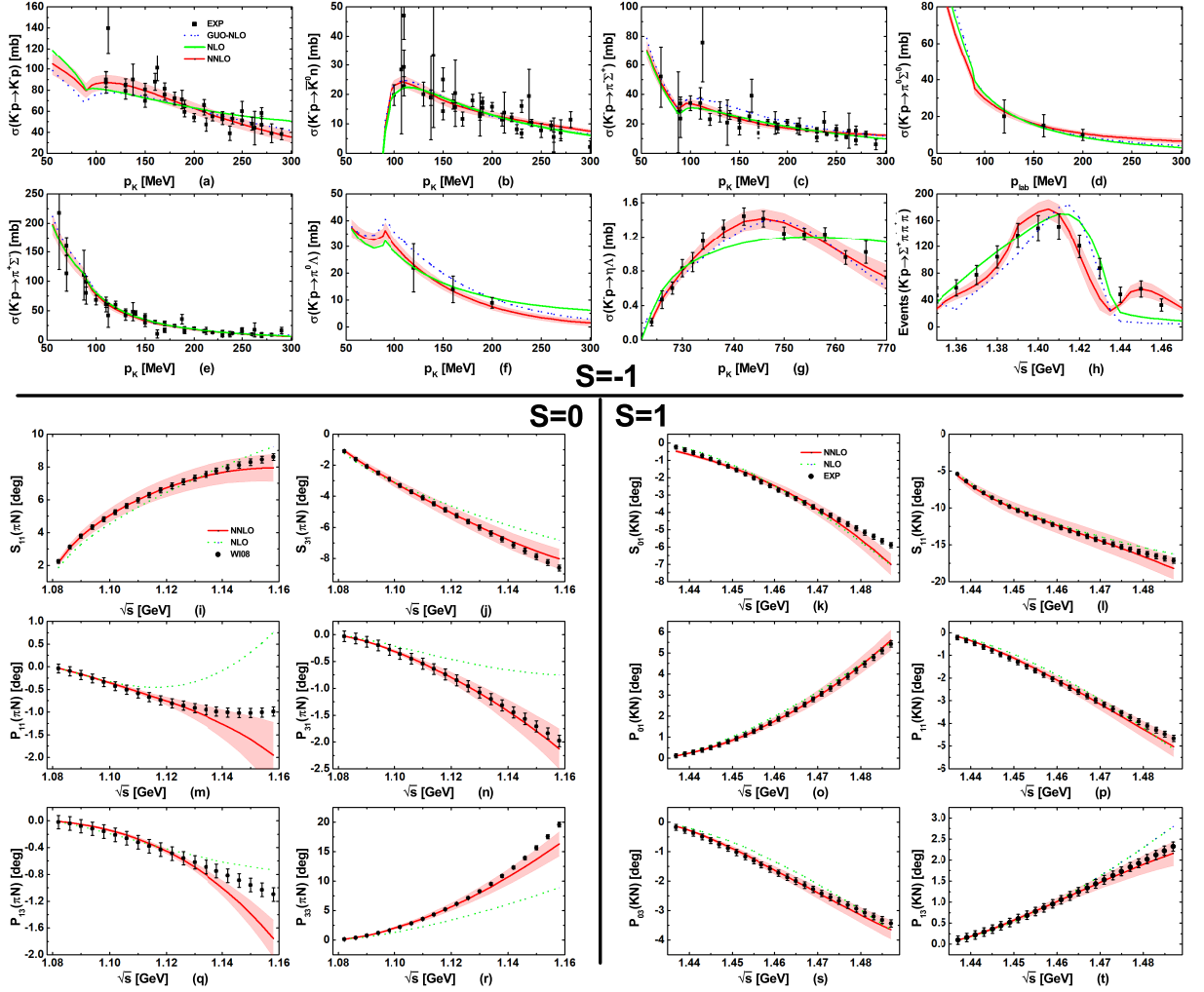}
\caption{Same as Fig.~\ref{crosssection:KbarN} of the main text but without constraints from baryon masses (Fit ``NNLO*'').}
\label{crosssection:KbarNNOMA}
\end{figure*}

\subsection{Two $I=1$ poles}

It is interesting to note that in our NNLO fit there exist two $I=1$ states around the $\bar{K}N$ threshold located at  $(1435,-39)$ MeV and $(1440,-135)$ MeV on the $(--++++)$ sheet, the order of which corresponds to $\pi\Lambda, \pi\Sigma, \bar K N, \eta\Lambda, \eta\Sigma, K\Xi$ respectively. Both states are well above the $K^-p$ threshold and appear as cusps on the real axis.
In the Fit ``NNLO*'' in which the constraints from baryon masses are omitted, the two $I=1$ states are located at $(1364,-110)$ MeV and $(1432,-18)$ MeV also on the $(--++++)$ sheet. In this case, the narrower state still shows up as a cusp but the broader one becomes a broad enhancement on the $I=1$ amplitude on the real axis.
We note that the existence of a $\Sigma^*(\frac{1}{2}^-)$ state has been predicted in a number of UChPT studies~\cite{Oller:2000fj,Khemchandani:2018amu} and also in the pentaquark model~\cite{Zhang:2004xt}. In  Refs.~\cite{Jido:2003cb,Roca:2013cca}, instead, a cusp located at the $\bar{K}N$ threshold was found, while in Ref.~\cite{Guo:2012vv}, two states were found, both of which are relatively narrow.  In addition, a  $\Sigma^*(\frac{1}{2}^-)$  state located at 1580 MeV was  found in Ref.~\cite{Jido:2003cb}. There is indeed some evidence for  the existence of a $\Sigma^*(\frac{1}{2}^-)$ state in the $K^-p\to \Lambda \pi^+\pi^-$ ~\cite{Wu:2009tu,Wu:2009nw} and ~$\gamma n\to K^+\Sigma^*(1385)$ reactions~\cite{Gao:2010hy}. Possible signals of this state have been studied in various other reactions~\cite{Zou:2006uh,Chen:2013vxa,Xie:2014zga,Wang:2015qta,Xie:2017xwx,Kim:2021wov}.  Clearly, the exact positions and the nature of these  states can  not yet be settled, calling for further experimental and theoretical studies.

\end{widetext}

\end{document}